\documentclass[12pt]{article}
\usepackage{amssymb}
\usepackage{amsmath, amsthm}
\newcommand{\be}{\begin{equation}}
\newcommand{\ee}{\end{equation}}
\begin{document}
\def\theequation{\arabic{section}.\arabic{equation}}
\begin{titlepage}
\title{Extension of the EGS theorem to metric and Palatini 
$f(R)$ gravity}
\author{Valerio Faraoni\\ \\
{\small \it Physics Department, Bishop's University}\\
{\small \it 2600 College St., Sherbrooke, Qu\'{e}bec, Canada 
J1M~1Z7}
}
\date{} \maketitle
\thispagestyle{empty}
\vspace*{1truecm}
\begin{abstract}
By using the equivalence between metric and Palatini $f(R)$ (or 
``modified'') gravities with $\omega=0, -3/2$ Brans-Dicke theories, it 
is shown that  the  Ehlers-Geren-Sachs theorem of general 
relativity is extended to 
modified gravity. In the case of  metric $f(R)$ 
gravity studied before, this agrees with previous literature.
 \end{abstract} \vspace*{1truecm}
\setcounter{page}{1}
\end{titlepage}



\section{Introduction}
\setcounter{equation}{0}
\setcounter{page}{2}

Through an exceptional number of publications on the subject, Sergei 
Odintsov has contributed to advance alternative 
theories of  gravity and cosmology  
motivated by quantum 
corrections to the  classical Einstein-Hilbert action. Among these are  
scalar-tensor and  $f(R)$  gravitational theories and, therefore,  it 
seems appropriate to include in this volume dedicated to Sergei a small 
contribution to cosmology in these theories.\footnote{A 
streamlined version of the discussion presented here was reported in 
\cite{review}.}

In relativistic cosmology, the identification of our universe 
with a Friedmann-Lemaitre-Robertson-Walker (FLRW) space relies 
on the 
observations of spatial homogeneity and isotropy around us. The 
strongest support for this assumption, which lies at the core of 
relativistic cosmology (in both Einstein's theory of general 
relativity  and in alternative gravitational theories) comes 
from the observation of the high degree of isotropy of the 
cosmic microwave background (CMB), supplemented by the assumption that 
isotropy would be observed from any spatial point in the 
universe (the Copernican principle --- such an assumption would be 
hard to check). The fact that a  
spacetime in which a family of  
observers exists who see the CMB isotropic around them can be identified 
with a FLRW space is far from trivial and, from the mathematical 
point of view, constitutes a kinematical characterization of 
FLRW spaces known as the Ehlers-Geren-Sachs (hereafter EGS) theorem 
\cite{EGS}.  Usually, the vanishing of acceleration, shear, and 
vorticity,
\be
\dot{u}^a\equiv u^c\nabla_c u^a=0 \;, \;\;\;\;\;
\sigma_{ab}=0 \;, \;\;\;\; \omega_{ab}=0 \;,
\ee
for a congruence of ``typical'' observers with four-velocity $u^a$ is 
taken to imply that the spacetime is of the FLRW type, {\em i.e.}, with 
line element
\be
ds^2=-dt^2+a^2(t) \gamma_{ij}(x^k) dx^i dx^j \;,
\ee
where $ \gamma_{ij}$ is a constant curvature 3-metric. However, this is 
guaranteed only if {\em i})~ matter is described by a perfect fluid, and 
{\em ii}) the Einstein equations are imposed. These conditions enforce   
the vanishing of the Weyl tensor.

In its original version, the EGS theorem states that, if a 
congruence of timelike, freely falling observers in a 
dust-dominated ({\em i.e.}, with vanishing pressure 
$P$) spacetime sees an isotropic radiation field, then 
(assuming that isotropy holds about every point) the spacetime 
is spatially homogeneous and isotropic and, therefore, a FLRW 
one. The original EGS theorem was generalized to an arbitrary 
perfect  fluid that is geodesic and barotropic  and 
with observers that are geodesics and irrotational \cite{Ellisetal, 
CBFMP}. 
Moreover, an ``almost EGS theorem'' has been proved: spacetimes 
that are close to satisfying the EGS conditions are close to 
FLRW spaces in an appropriate sense\footnote{For a discussion of 
inhomogeneous or anisotropic cosmological models admitting an isotropic 
radiation field, see \cite{Clarkson2}.} \cite{Stoegeretal, 
 Taylor:1995dy}. Perhaps it is not exaggerated to regard  
the 
EGS theorem and its 
generalizations  as a cornerstone of relativistic cosmology motivating 
the use of the standard Big Bang model.

Since the discovery of the EGS theorem in 1968, theoretical 
cosmology has expanded considerably to include the possibility that 
general relativity  
may 
have to be augmented  by adding quantum corrections which take the form 
of  extra terms in the 
Einstein-Hilbert action\footnote{Here $R$ is the Ricci curvature 
of the  metric $g_{ab}$, which  has determinant $g$ and $\kappa 
\equiv 8\pi G$, ~$G$ is Newton's constant, and $S^m$ denotes the 
matter part of the action. We follow the notations of 
Ref.~\cite{Wald}.} $ S_{EH}=\frac{1}{2\kappa}\int d^4 
x \, 
\sqrt{-g} \, R+S^m $. Here we 
are interested in theories described by an action of the form
\be \label{actionmetric}
S=\frac{1}{2\kappa}\int d^4 x \, \sqrt{-g} \, f(R)+S^m \;,
\ee
where $f(R) $ is an arbitrary (but twice differentiable) 
function  of its argument. While $f(R)$ gravity has a long 
history 
(\cite{Weyl} - see \cite{Schmidt} for an 
historical review), quadratic corrections to general relativity , {\em 
i.e.},  
$f(R)=R+\alpha R^2$, were introduced early on  as 
semiclassical corrections or counterterms to renormalize 
general relativity \cite{renorma}: such corrections, which are motivated 
also by 
string theories \cite{stringmotivations}, are obviously 
important at large curvatures, {\em e.g.}, in the very early 
universe (in  which they have been used to propel inflation in 
Starobinski's scenario \cite{Starobinsky}) and near black holes, 
in which non-linear choices of the function $f(R)$ may cure the 
problem of the central singularity \cite{blackholes}. More 
recently, 
``modified'' or ``$f(R)$'' gravity has seen a new lease on life 
after the 1998 discovery of the acceleration of the cosmic 
expansion using type~Ia supernovae \cite{SN}. While one 
possibility is to explain the present acceleration of the 
universe by postulating  a mysterious form of dark energy with 
exotic properties ($P\simeq -\rho$, where $\rho$ is the energy 
density), it has been proposed that perhaps we are seeing the 
first deviations from Einstein's gravity on very large scales. 
The prototypical modification of the Einstein-Hilbert action 
consisted of the choice $f(R)=R-\mu^4/R$, where $\mu \simeq 
H_0^{-1}$ is a mass scale of the order of the present value of 
the Hubble parameter, {\em i.e.,} extremely small on particle 
physics scales \cite{CCT, CDTT}. While this particular model is 
in gross 
violation of the Solar System observational constraints on 
the parametrized-post-Newtonian parameter $\gamma$ 
\cite{CSE} 
and is subject to a violent instability \cite{DolgovKawasaki, 
Odintsovconfirm, mattmodgrav}, choices of the function $f(R)$  
that satisfy the experimental constraints and provide the 
correct cosmological dynamics abound in the literature 
(\cite{designerf(R)} --- see
 \cite{NojiriOdintsovreview, review} for reviews). Three versions of 
modified gravity exist: the first is {\em metric $f(R)$ gravity}, in 
which 
the action~(\ref{actionmetric}) is varied with respect to the 
metric, and provides the fourth order field equations
\be \label{metricfe}
 f'(R)R_{ab}-\frac{1}{2}f(R)g_{ab}- 
\left[\nabla_a \nabla_b -g_{ab}\Box\right] f'(R)= 
\kappa \,T_{ab}, 
\ee
where, as usual, $
T_{ab}=\frac{-2}{\sqrt{-g}}\, \frac{\delta
S_M }{\delta g^{ab} }  $, and 
a prime denotes differentiation with respect to $R$.

The second version of the theory is {\em Palatini $f(R)$ gravity}, in 
which the metric $g_{ab}$ and 
the connection $\Gamma^a_{bc}$ are considered as independent 
quantities ({\em i.e.}, the connection is not identified with 
the metric connection $\left\{^a_{bc} \right\}$ of $g_{ab}$), the 
Ricci tensor ${\cal R}_{ab}$ is constructed out this non-metric 
connection, and ${\cal R}\equiv g^{ab}{\cal R}_{ab}$ \cite{Vollick}. The 
Palatini action  is 
\be \label{actionPalatini}
S=\frac{1}{2\kappa}\int d^4 x \, \sqrt{-g} \, f( {\cal R})+S^m 
\;,
\ee
in which the matter part of the action does not depend 
explicitly on the (non-metric) connection. 
Variation with respect to the metric and the connection 
$\Gamma^a_{bc}$ provides the second order field equations
\begin{eqnarray} 
f'({\cal R}) {\cal R}_{(ab)}-\frac{1}{2}f({\cal 
R})g_{ab}&=&\kappa \,G\, T_{ab} \; , \label{Palatinife1}\\
&& \nonumber \\
\bar{\nabla}_c\left(\sqrt{-g}f'({\cal R})g^{ab}\right)&=&0 \; 
,
\label{Palatinife2}
\end{eqnarray}
respectively, where $\tilde{\nabla}_a$ denotes the covariant derivative 
operator of $ \Gamma^a_{bc}$.  Little is still known about a  third 
version ({\em 
metric-affine gravity}), 
in which the matter action is allowed to depend explicitly on 
the (non-metric) connection \cite{metricaffine}, and which will not be 
considered here.

\section{The EGS theorem in $f(R)$ gravity}
\setcounter{equation}{0}

Since the possibility that our present universe is described 
by some  modification of general relativity is now  taken 
rather seriously, and $f(R)$ gravity is at least a convenient toy model 
(if not a serious candidate), it is natural to ask whether the basic 
result 
that allows one to  identify the observed universe with a FLRW space, 
the EGS  theorem, survives in  these theories. The first 
investigations \cite{Maartens:1994pb, Taylor:1995dy} gave an 
affirmative answer for the theory
\be \label{Roy}
S=\frac{1}{2\kappa}\int d^4 x \, \sqrt{-g} \left(  R +\alpha 
R^2+\beta R_{ab}R^{ab} \right)+S^m 
\ee
in the metric formalism,\footnote{Due to the fact that the Gauss-Bonnet 
expression $R^2-4R_{ab}R^{ab}+R_{abcd}R^{abcd}$ gives rise to a 
topological invariant, it is unnecessary to include Riemann-squared 
terms in the action~(\ref{Roy}).} subject to the additional condition 
(which 
does not appear in general relativity) that the perfect fluid filling 
the universe obeys a barotropic equation of state $P=P(\rho)$ with 
$dP/d\rho \neq 0$, which implies that surfaces of constant $P$ and 
surfaces of constant $\rho$ coincide. Subsequently, the validity of 
the EGS theorem was extended to general {\em metric} $f(R)$ gravity in 
Ref.~\cite{Rippl:1995bg}. Here we extend this result to {\em 
Palatini} modified gravity, and we provide an independent  proof 
also for the metric verson of these 
theories. The result is straightforward because it builds on the 
results of Ref.~\cite{ClarksonColeyONeill} that extend the 
validity of the EGS theorem to scalar-tensor theories of gravity 
described by the action \cite{ST}
\be 
S_{ST}=\frac{1}{2\kappa}\int d^4 x \, \sqrt{-g} \left[ \psi R 
-\frac{\omega(\psi )}{\psi} \, \nabla^c \psi \nabla_c \psi 
-V(\psi) \right] +S^m \;,
\ee
where the coupling function $\omega (\psi)$ generalizes the 
constant Brans-Dicke parameter \cite{BD}. Now, it is 
well-known that metric or Palatini $f(R)$ gravity can be seen as 
a Brans-Dicke theory \cite{STequivalence, Wands}. 
In the metric formalism, with the introduction of an extra field 
$\phi$, the action (\ref{actionmetric}) can be rewritten 
as 
\begin{equation} \label{STaction}
S=\frac{1}{2\kappa} \int d^4 x \,  \sqrt{-g}\,\left[ \psi(\phi) 
R-V(\phi) \right] +S^{(m)} 
\end{equation}
when   
$f''(R) \neq 0$, where $\phi$ is defined by
\begin{equation} \label{1400} 
\psi(\phi)=f'(\phi) 
\ee
and 
\be 
 V(\phi)=\phi f'(\phi) -f(\phi) \;.
\end{equation}
The action~(\ref{STaction}) reduces to  (\ref{actionmetric})  
trivially if $\phi=R$ and, vice-versa, variation 
of~(\ref{STaction}) with respect to $\phi$ yields 
\be
 \left( 
R-\phi \right) f''(R)=0 \;,
\ee
which implies that $\phi=R$  if 
$f''\neq 0$. The action can now be seen as a Brans-Dicke action 
with Brans-Dicke parameter $\omega=0$ 
if the field  $\psi \equiv f'(\phi)=f'(R)$ is used instead of 
$\phi$ as the 
independent Brans-Dicke-like field:
\be \label{BD}
S=\frac{1}{2\kappa} \int d^4x \sqrt{-g} \left[ \psi R -U( 
\psi) \right] +S^{(m)} \;,
\ee
where 
\be
 U(\psi)=V\left( \phi(\psi) \right) =\psi 
\phi(\psi)-f\left( \phi(\psi) \right) 
\ee
(this is called ``O'Hanlon theory'' or  ``massive dilaton gravity''  
\cite{O'Hanlon, Wands}).

In the Palatini formalism,  by introducing the metric 
$h_{ab}\equiv f'(\tilde{R}) \, g_{ab} $ conformally 
related to $g_{ab}$ and the scalar $ \phi \equiv f'\left( 
{\cal R} \right)$, and using the 
transformation  property of the Ricci scalar 
\cite{Wald}, one obtains
\be\label{Riccitransform}
{\cal R}=R +\frac{3}{2\phi} \, \nabla^c\phi\nabla_c\phi 
-\frac{3}{2}\, \Box \phi \;,
\ee
and the action  is equivalent to
\be
S=\frac{1}{2\kappa} \int d^4 x \, \sqrt{-g} \left[ f(\chi) 
+f'(\chi)\left( \cal{R}-\chi \right)\right] +S^{(m)}
\ee
if $f''\neq 0$ with $\chi=\tilde{R}$. By redefining $\chi$ 
through $ \phi=f'\left( \chi \right)$, it is 
\be
f(\chi)+f'(\chi)\left( {\cal R}-\chi \right)=\phi 
{\cal R}-\phi \chi (\phi) +f\left( \chi(\phi) \right)=
\phi {\cal R} +\frac{3}{2}\, \nabla^c\phi\nabla_c\phi 
-V(\phi)-3\Box 
\phi 
\ee
using eq.~(\ref{Riccitransform}), where
\be
V(\phi)=\phi\chi(\phi)-f(\chi(\phi)) \;.
\ee
Apart from a boundary term, this yields
\be \label{STequivalentPalatini}
S=\frac{1}{2\kappa}\int d^4x \, \sqrt{-g} \left[ \phi {\cal R} 
+\frac{3}{2 \phi}\, 
\nabla^c\phi\nabla_c \phi -V(\phi) 
\right] +S^{(m)}
\ee
which describes a Brans-Dicke theory with parameter  
$\omega=-3/2 $, seldom considered in the literature \cite{ST-3/2} until 
the recent attempts to model the present-day cosmic acceleration.

Now, since the EGS theorem has been proved to hold  for 
scalar-tensor gravity 
\cite{ClarksonColeyONeill}, it is straightforward to conclude 
that its validity is extended to metric modified gravity. This quick 
proof is 
consistent with the results obtained in   
\cite{Rippl:1995bg} for 
general metric $f(R)$ gravity with a more direct approach, and with the  
findings of \cite{Maartens:1994pb, Taylor:1995dy} for quadratic $f(R)$.  
The validity of the EGS theorem is then extended 
to Palatini $f(R)$ gravity, which was not considered before in the 
EGS context. This is not entirely trivial when one considers  
the different order 
of the field equations  with respect 
to metric $f(R) $ theories (second order instead of fourth),  and the 
different physics described.

\section{Conclusions}
\setcounter{equation}{0}

The equivalence between metric and Palatini $f(R)$ theories  and 
$\omega=0, -3/2$ Brans-Dicke theories allows for a 
straightforward proof of the EGS theorem for modified  
gravity, which relies on the previous work  
\cite{ClarksonColeyONeill} extending the validity of this theorem to 
scalar-tensor gravity. By contrast, the direct approach of 
\cite{Rippl:1995bg} appears a bit cumbersome. 

In addition to providing  a different approach to the EGS theorem for 
metric $f(R)$ gravity, we provide  a straightforward proof of its  
validity for {\em Palatini} $f(R)$ gravity, which was not considered 
before in this context. Although evidence is now accumulating that 
Palatini  modified gravity is not physically viable for various reasons 
(see  \cite{BarausseSotiriouMiller, TremblayFaraoni, 
PalatiniPLB, FaraoniTremblay, 
review}  for a discussion of the various theoretical aspects 
involved), it may still be useful as a toy model to analyze  
mathematical and physical features of generalized gravity theories.

Given the fact that the EGS theorem extends to Lagrangian 
densities of the form ${\cal L}= R+\alpha 
R^2+\beta R_{ab}R^{ab} +{\cal L}^m$ \cite{Maartens:1994pb, 
Taylor:1995dy}, 
one wonders whether it is actually valid for more general theories of 
the form $ {\cal L} = f\left( R, R_{ab}R^{ab}, R_{abcd}R^{abcd} 
\right)$, which are motivated by low-energy string corrections 
to general relativity and are not equivalent to a  simple 
scalar-tensor 
theory.  This possibility will be examined elsewhere.

\section*{Acknowledgments}

It is a pleasure to  acknowledge several discussions on scalar-tensor 
and $f(R)$ gravity  with 
Sergei Odintsov over the years. This work is supported by the Natural 
Sciences and Engineering Research Council of Canada  (NSERC).

\end{document}